\documentclass[%
 reprint,
 amsmath,amssymb,
 aps,
]{revtex4-1}

\usepackage{graphicx}
\usepackage{dcolumn}
\usepackage{bm}

\begin{document}

\preprint{APS/123-QED}

\title{Improving student understanding of addition of \\
angular momentum in
quantum mechanics}

\author{Guangtian Zhu$^{1,2}$}
\author{Chandralekha Singh$^2$}
\affiliation{$^1$School of Education Science, East China Normal University, Shanghai,
China, 200062}
%
\affiliation{%
$^2$Department of Physics and Astronomy, University of Pittsburgh,
Pittsburgh, PA, 15260, USA\\
}%

%
%

\date{\today}

\begin{abstract}
We describe the difficulties advanced undergraduate and graduate students
have with concepts related to addition of angular momentum in quantum
mechanics. We also describe the development and implementation of a
research-based learning tool, a Quantum Interactive Learning Tutorial
(QuILT), to reduce these difficulties. The preliminary evaluation shows that
the QuILT related to the basics of the addition of angular momentum is
helpful in improving students' understanding of these concepts.
\end{abstract}

\pacs{Valid PACS appear here}
\maketitle


\section{\label{sec:level1}introduction}

Quantum mechanics is a particularly challenging subject for undergraduate
students. Based upon the research studies that have identified
difficulties \cite{1,2,3,4,5,6,7,8,9}, we have developed a set of research-based learning
tools to help students develop a good grasp of quantum mechanics \cite{10,11,12,13,14}.
These research-based learning tools include the Quantum Interactive Learning
Tutorials (QuILTs) and concept tests similar to those popularized by Mazur
for introductory physics courses \cite{15}. The QuILTs use a guided
inquiry-based approach to learning and help students in building a knowledge
structure by guiding them to discern the structure of quantum mechanics. The
instructors can use the QuILTs as either in-class tutorials or homework
supplements \cite{12,13}. The concept tests are integrated with lectures and
encourage students to take advantage of their peers' expertise and learn
from each other \cite{15}.

In this paper, we focus on our investigation to identify student
difficulties with concepts related to the addition of angular momentum in
quantum mechanics. In the course of this investigation, we found that the
main source of difficulties with concepts related to the addition of angular
momentum was that the students were not comfortable with the pre-requisites,
e.g., the basics of a single spin system (dimensionality of a vector space,
how to choose a basis and write operators in a given basis, etc.) and the
basics of multi-spin systems (dimensionality of a product space, how to
write a complete set of basis vectors in the product space, e.g., in the
coupled or uncoupled representation, how to write operators in a given
basis, etc.). Then, we developed and assessed a research-based QuILT to help
undergraduate students better grasp the preliminary concepts to aid them in
understanding the formalism of the addition of angular momentum. The QuILT
can also be used by underprepared graduate students as a self-study tool.
The investigation of students' difficulties with the addition of angular
momentum was conducted with the undergraduate and graduate students at the
University of Pittsburgh (Pitt) and other universities \cite{16,17,18,19} by
administering written tests and by conducting in-depth individual interviews
with a subset of them.

The research-based QuILT relating to the pre-requisites for the addition of
angular momentum was administered to students in the second semester of a
full-year junior-senior level quantum mechanics course. It strives to build
on students' prior knowledge, actively engaging them in the learning process
and helping them build connections between the abstract formalism and
conceptual aspects of quantum physics, without compromising the technical
content. To assess the effectiveness of the QuILT, a pre-test and a
post-test related to the addition of angular momentum were given to two
classes of undergraduate students at Pitt. We will discuss the results and
findings in a later section.

\section{Background}

Classically, the angular momentum vector $\vec {L}$ is defined by the cross
product of the position $\vec {r}$ and momentum $\vec {p}$, i.e., $\vec
{L}=\vec {r}\times \vec {p}$. In quantum mechanics, in which for every
observable there is an operator, the components of the orbital angular
momentum operator, $\hat {L}_x $, $\hat {L}_y $, and $\hat {L}_z $, do not
commute with each other ($[\hat {L}_x, \hat {L}_y ]=i\hbar \hat {L}_z
$, $[\hat {L}_y, \hat {L}_z ]=i\hbar \hat {L}_x $, $[\hat {L}_z, \hat {L}_x
]=i\hbar \hat {L}_y )$ and therefore the components of orbital angular
momentum are mutually incompatible observables. The eigenvalues of the
square of the magnitude of the orbital angular momentum operator, $\hat
{L}^2$, are $\ell (\ell +1)\hbar ^2$, where $\ell $, the orbital angular
momentum quantum number, is a non-negative integer and $\hbar \equiv h/(2\pi
)$ is the reduced Planck's constant. The eigenvalues of $\hat {L}_z $ are
$m\hbar $, where $m=-\ell, -\ell +1,\ldots,\ell $. Since $\hat {L}^2$ and $\hat
{L}_z $ commute ($[\hat {L}^2,\hat {L}_z ]=0$), we can use the orbital
angular momentum quantum numbers $\ell $ and $m$ to denote their simultaneous
eigenstates as $\left| {\ell, m} \right\rangle $.

In addition to the orbital angular momentum, $\vec {L}$, elementary
particles, such as electrons, also possess intrinsic spin angular momentum,
$\vec {S}$, which is not due to motion in position space. The algebras of
the orbital and spin angular momenta are similar and the components of the
spin angular momentum operator, $\hat {S}_x $, $\hat {S}_y $, and $\hat
{S}_z $, satisfy commutation relations similar to the commutation relations
among the components of the orbital angular momentum operator, $\hat {L}_x
$, $\hat {L}_y $ and $\hat {L}_z $, {\it i.e.}\ $[\hat {S}_x, \hat {S}_y ]=i\hbar
\hat {S}_z $, $[\hat {S}_y, \hat {S}_z ]=i\hbar \hat {S}_x $, $[\hat {S}_z
,\hat {S}_x ]=i\hbar \hat {S}_y $, $[\hat {S}^2,\hat {S}_z ]=0$. The
eigenvalues of the square of the magnitude of the spin angular momentum
operator, $\hat {S}^2$, are $s(s+1)\hbar ^2$, where $s$ is the spin quantum
number. The spin quantum number, $s$, can be a non-negative integer or a
non-negative half-odd-integer. For the electron, the spin quantum number,
$s$, is $1/2$ and the values of $m_s $, the spin quantum number for the
$z$-component of spin, are $\pm 1/2$. If we choose the eigenstates of the
$z$-component of spin as the basis vectors, the operators, $\hat {S}_x
=\frac{\hbar }{2}\hat {\sigma }_x $, $\hat {S}_y =\frac{\hbar }{2}\hat
{\sigma }_y $, and $\hat {S}_z =\frac{\hbar }{2}\hat {\sigma }_z $, can be
represented by the Pauli matrices, $\hat {\sigma }_x =\left(
{{\begin{array}{*{20}c}
 0 \hfill & 1 \hfill \\
 1 \hfill & 0 \hfill \\
\end{array} }} \right)$, $\hat {\sigma }_y =\left( {{\begin{array}{*{20}c}
 0 \hfill & {-i} \hfill \\
 i \hfill & 0 \hfill \\
\end{array} }} \right)$, and $\hat {\sigma }_z =\left(
{{\begin{array}{*{20}c}
 1 \hfill & 0 \hfill \\
 0 \hfill & {-1} \hfill \\
\end{array} }} \right)$, respectively. Because $\hat {S}^2$ and $\hat {S}_z
$ commute, we can use the quantum numbers $s$ and $m_s $ to denote
their simultaneous eigenstates as $\left| {s,m_s } \right\rangle $.

If a quantum system contains two particles with individual orbital angular
momentum quantum numbers $\ell _1 $ and $\ell _2 $, the total orbital
angular momentum quantum number of the system can range from $\ell _1 +\ell
_2 $ down to $\left| {\ell _1 -\ell _2 } \right|$, i.e., $\ell =\ell _1
+\ell _2 $, $\ell _1 +\ell _2 -1$, {\ldots}, $\left| {\ell _1 -\ell _2 }
\right|$. The $z$-component of the total orbital angular momentum of the system
equals the sum of the $z$-components of the orbital angular momenta of the
individual particles, i.e., $m=m_1 +m_2 $. For a single particle with
non-zero spin, the possible values of its total angular momentum quantum
number, $j$, can be obtained by the addition of its orbital angular momentum
quantum number, $\ell $, and its spin angular momentum quantum number, $s$,
appropriately, i.e., $j=\ell +s$, $\ell +s-1$, {\ldots}, $\left| {\ell -s}
\right|$. Similarly, for two particles with total angular momentum quantum
numbers $j_1 $ and $j_2 $, the total angular momentum quantum number of the
system is $j=j_1 +j_2 $, $j_1 +j_2 -1$, {\ldots}, $\left| {j_1 -j_2 }
\right|$.

In the junior/senior level quantum mechanics courses, the QuILT is developed
with learning goals to help students develop a better understanding of the
following three issues related to addition of angular momentum.

\subsection{Recognizing the dimension of a Hilbert space}

The dimension of a Hilbert space is equal to the number of linearly
independent basis vectors, e.g., the number of linearly independent system
eigenstates of any operator representing a system observable that acts on
the states in that space. For example, for a particle in a one-dimensional
(1D) infinite square well, the infinitely many energy eigenstates $\left|
{\psi _n } \right\rangle $ of the Hamiltonian operator form a complete set
of basis vectors for the infinite dimensional Hilbert space.

The Hilbert space corresponding to the spin angular momentum of a single
spin-1/2 particle is two dimensional. For example, the $z$-component of the
spin of an electron has two eigenstates, $\left| {s=1/2,m_s =1/2}
\right\rangle $ and $\left| {s=1/2,m_s =-1/2} \right\rangle $ (or $\left|
{m_s =1/2} \right\rangle $ and $\left| {m_s =-1/2} \right\rangle $ for
short, since $s=1/2$ is a fixed number for an electron). If a system
consists of two electrons, the product space corresponding to the spin
degrees of freedom will be four dimensional, which is the product of the
dimensions of the Hilbert spaces of each of the spins separately. The basis
vectors of the four dimensional product space in the uncoupled
representation are $\left| {m_{s1} =1/2} \right\rangle \otimes \left|
{m_{s2} =1/2} \right\rangle $, $\left| {m_{s1} =1/2} \right\rangle \otimes
\left| {m_{s2} =-1/2} \right\rangle $, $\left| {m_{s1} =-1/2} \right\rangle
\otimes \left| {m_{s2} =1/2} \right\rangle $ and $\left| {m_{s1} =-1/2}
\right\rangle \otimes \left| {m_{s2} =-1/2} \right\rangle $ (``$\otimes $''
is used to denote the direct product or the Kronecker product \cite{20}).

\subsection{Choosing the basis vectors for the Hilbert space}

For a system consisting of two spin-1/2 particles, there are two common ways
to represent the basis vectors for the product space. Since the spin quantum
numbers $s_1 =1/2$ and $s_2 =1/2$ are fixed, we can use the ``uncoupled
representation'' and express the orthonormal basis vectors for the product
space as $\left| {s_1, m_1 } \right\rangle \otimes \left| {s_2, m_2 }
\right\rangle =\left| {m_1 } \right\rangle \otimes \left| {m_2 }
\right\rangle $, as noted earlier. In this uncoupled representation, the
operators related to each particle (subspace) act on their own states, e.g.,
$\hat {S}_{1z} \left| {1/2} \right\rangle _1 \otimes \left| {-1/2}
\right\rangle _2 =\frac{\hbar }{2}\left| {1/2} \right\rangle _1 \otimes
\left| {-1/2} \right\rangle _2 $ and $\hat {S}_{2z} \left| {1/2}
\right\rangle _1 \otimes \left| {-1/2} \right\rangle _2 =-\frac{\hbar
}{2}\left| {1/2} \right\rangle _1 \otimes \left| {-1/2} \right\rangle _2 $.
On the other hand, we can use the ``coupled representation'' and find the
total spin quantum number for the system of two particles together. The
total spin quantum number for the two spin-1/2 particle system, $s$, is either
$1/2+1/2=1$ or $1/2-1/2=0$. When the total spin quantum number $s$ is 1, the
quantum number $m_s $ for the $z$-component of the total spin, $S_z $, can be
1, 0, and $-$1. When the total spin is 0, $m_s $ can only be 0. Therefore, the
basis vectors of the system in the coupled representation are $\left|
{s=1,m_s =1} \right\rangle $, $\left| {s=1,m_s =0} \right\rangle $, $\left|
{s=1,m_s =-1} \right\rangle $ and $\left| {s=0,m_s =0} \right\rangle $. In
the coupled representation, the state of a two-spin system is not a simple
product of the states of each individual spin although we can write each
coupled state as a linear superposition of a complete set of uncoupled
states. For example, the normalized basis vectors in the coupled representation $\left|
{s=1,m_s =1} \right\rangle $, $\left| {s=1,m_s =0} \right\rangle $, $\left|
{s=1,m_s =-1} \right\rangle $ and $\left| {s=0,m_s =0} \right\rangle $ can
be expressed in the uncoupled representation as follows:
\begin{eqnarray*}
\left| {s=1,m_s =1} \right\rangle &=&\left| {m_1 =1/2} \right\rangle \otimes
\left| {m_2 =1/2} \right\rangle,\\
\left| {s=1,m_s =0} \right\rangle &=&(\left| {m_1 =1/2} \right\rangle \otimes
\left| {m_2 =-1/2} \right\rangle \\
&& +\left| {m_1 =-1/2} \right\rangle \otimes
\left| {m_2 =1/2} \right\rangle )/\sqrt 2,\\
\left| {s=1,m_s =-1} \right\rangle &=&\left| {m_1 =-1/2} \right\rangle \otimes
\left| {m_2 =-1/2} \right\rangle,\\
\left| {s=0,m_s =0} \right\rangle &=&(\left| {m_1 =1/2} \right\rangle \otimes
\left| {m_2 =-1/2} \right\rangle \\
&& -\left| {m_1 =-1/2} \right\rangle \otimes
\left| {m_2 =1/2} \right\rangle )/\sqrt 2.
\end{eqnarray*}

\subsection{Constructing the matrix of angular momentum operators}

To calculate the diagonal and off-diagonal matrix elements of an operator in
the product space, we must compute the matrix elements of that operator in
the appropriate basis. For example, for a two spin-1/2 particle system, for
the operator $\hat {S}_{1z} +\hat {S}_{2z} $, when we use the basis vectors
in the uncoupled representation, the matrix elements are $\left\langle
{{m}'_2 } \right|\otimes \left\langle {{m}'_1 } \right|\hat {S}_{1z} +\hat
{S}_{2z} \left| {m_1 } \right\rangle \otimes \left| {m_2 } \right\rangle $
where $m_1 $, $m_2 $, ${m}'_1 $, ${m}'_2 $ are either 1/2 or $-$1/2. If we
choose the order of the basis vectors to be $\left| {1/2} \right\rangle _1
\otimes \left| {1/2} \right\rangle _2 $, $\left| {1/2} \right\rangle _1
\otimes \left| {-1/2} \right\rangle _2 $, $\left| {-1/2} \right\rangle _1
\otimes \left| {1/2} \right\rangle _2 $, and $\left| {-1/2} \right\rangle _1
\otimes \left| {-1/2} \right\rangle _2 $, the operator matrix is
\[
\hat {S}_{1z} +\hat {S}_{2z} =\left( {{\begin{array}{*{20}c}
 \hbar \hfill & 0 \hfill & 0 \hfill & 0 \hfill \\
 0 \hfill & 0 \hfill & 0 \hfill & 0 \hfill \\
 0 \hfill & 0 \hfill & 0 \hfill & 0 \hfill \\
 0 \hfill & 0 \hfill & 0 \hfill & {-\hbar } \hfill \\
\end{array} }} \right).
\]
The basis vectors in the coupled representation $\left| {s,m_s }
\right\rangle $ are also a good choice to express this operator in matrix
form and the matrix is diagonal since $\left| {s,m_s } \right\rangle $ are
the eigenstates of the $z$-component of the total spin operator $\hat {S}_z
=\hat {S}_{1z} +\hat {S}_{2z} $. When we construct the operator matrix by
using the basis vectors for the coupled representation in the order $\left|
{1,1} \right\rangle $, $\left| {1,0} \right\rangle $, $\left| {1,-1}
\right\rangle $, and $\left| {0,0} \right\rangle $, the operator matrix can
be expressed as
\[
\hat {S}_{1z} +\hat {S}_{2z} =\left( {{\begin{array}{*{20}c}
 \hbar \hfill & 0 \hfill & 0 \hfill & 0 \hfill \\
 0 \hfill & 0 \hfill & 0 \hfill & 0 \hfill \\
 0 \hfill & 0 \hfill & {-\hbar } \hfill & 0 \hfill \\
 0 \hfill & 0 \hfill & 0 \hfill & 0 \hfill \\
\end{array} }} \right).
\]
We can also rearrange the sequence of the basis as $\left| {1,1}
\right\rangle $, $\left| {1,-1} \right\rangle $, $\left| {1,0} \right\rangle
$ and $\left| {0,0} \right\rangle $ in order to move the non-zero matrix
elements, e.g., to the upper left corner. Although the matrices of the
particular operator $\hat {S}_z =\hat {S}_{1z} +\hat {S}_{2z} $ are diagonal in
both the coupled and uncoupled representations, in general, the operator matrices
are different in different basis sets (an issue discussed at length in the
QuILT). In particular, the matrix corresponding to an operator may be diagonal in one
representation but not in another representation.

\section{Investigation of students' difficulties}

The investigation of difficulties was carried out by administering
free-response and multiple-choice questions to advanced undergraduate
students enrolled in quantum mechanics courses. Individual interviews were
also carried out, to better understand students' rationale for their
responses, before, during and after the development of different versions of
the QuILT on the preliminaries of the addition of angular momentum and the
corresponding pre-test and post-test. In addition to informal discussions
with students, we conducted one-on-one interviews (each lasting between 1--2
hours) with 15 undergraduate and graduate students using a think-aloud
protocol. In the interviews, students were asked to articulate their
reasoning processes while they answered the questions. Students were not
interrupted unless they remained silent for a while. At the end of the
interviews, students were asked to clarify the issues they had not made
clear earlier. After each individual interview on a particular version of
the QuILT (along with the pre-test and the post-test administered),
modifications were made based upon the feedback obtained from students'
performance on the QuILT, the pre-test and the post-test. For example, if
students got stuck at a particular point and could not make progress from
one question to the next, modifications were made accordingly.

\subsection{Difficulty with the dimension of a Hilbert space}

\noindent\textbf{\textit{Difficulty A.1: Confusion between Hilbert space and position
space even }}\underline {\textbf{\textit{before}}}\textbf{\textit{ learning
about addition of angular momentum }}
\vskip 0.5em
We have found that the concepts related to Hilbert space are very difficult
for students and many students were confused about the dimensions of the
Hilbert space and the position space. The following multiple choice question
was given to 33 undergraduate students (before instruction in addition of
angular momentum, but after instruction in relevant concepts) to probe
whether they could distinguish between the one-dimensional position space in
which the particle is confined and the infinite-dimensional Hilbert space of
system states.

\begin{enumerate}
\item[\textbullet] \textit{Choose all of the following statements that are correct for a particle interacting with a one dimensional (1D) infinite square well.}
	\item[] \textit{(1) The appropriate Hilbert space for this system is one dimensional.}
	\item[] \textit{(2) The energy eigenstates of the system form a basis in a 1D Hilbert space.}
	\item[] \textit{(3) The position eigenstates of the system form a basis in a 1D Hilbert space.}
\item[] \textit{A. none of the above B. 1 only C. 2 only D. 3 only E. all of the above}
\end{enumerate}

The Hilbert space in which the state of the system lies is infinite
dimensional while the position space in which the particle is confined is
one dimensional. However, only 48{\%} of the students chose the correct
answer A. About 25{\%} of the students selected the answer E and incorrectly
believed that both the energy eigenstates and position eigenstates form a
basis in a 1D Hilbert space. This example illustrates that the students have
difficulties with the dimension of the Hilbert space even before addition of
angular momentum is discussed in the quantum mechanics course. For a given
quantum system, the translational and spin degrees of freedom are distinct.
For example, the vector space corresponding to the spin angular momentum or
the orbital angular momentum of an atom is finite dimensional, while the
vector space of the translational degrees of freedom is infinite
dimensional. Therefore, when discussing the angular momentum of quantum
particles, the translational degrees of freedom of the system were not
emphasized in the QuILT in order to maintain students' focus on addition of angular
momentum and avoid further confusions about the dimension of Hilbert space.
\vskip 0.5em
\noindent\textbf{\textit{Difficulty A.2: Incorrectly calculating the dimension of a
product space by adding the dimensions of the subspaces}}
\vskip 0.5em
Students in general have great difficulty finding the dimension of a product
space containing two or more angular momenta. When asked about the dimension
$D$ of a product space consisting of two subspaces of dimensions $D_1 $ and
$D_2 $, many students incorrectly believed that $D=D_1 +D_2 $ instead of
$D_1 \times D_2 $. Discussions with individual students suggest that such a
misconception often originates from two reasons. One reason is the word
``addition'' in ``addition of angular momentum''. The second reason is
related to the simplest example in which students learn about the product
space for two spin-1/2 particles. In this case, the dimension of the product
space is four, which equals $2\times 2$ but is also $2+2$. When we asked 11
students about the dimension of the product space for a system containing
two spin-1 particles, 4 of them provided the incorrect answer $6=3+3$
instead of the correct answer $9=3\times 3$. Individual discussions with
students also suggest that students are confused about the dimension of the
product space. It appears, therefore, that using the example of two
two-dimensional spaces is a poor choice unless students are also given ample
opportunity to contemplate the results for other product spaces (which is
done in the QuILT). Due to its simplicity, the product space for two
two-dimensional spaces is the choice commonly used to illustrate issues
related to the addition of angular momentum and students may or may not have
the opportunity to extend these results to other cases.

\subsection{Difficulty in identifying different basis vectors for the product space}

\noindent\textbf{\textit{Difficulty B.1: Difficulty in choosing a convenient basis to
represent an operator as an }}$N\times N$\textbf{\textit{matrix in an
N-dimensional product space}}
\vskip 0.5em
Students often have difficulty in figuring out when it would be convenient
to choose the basis vectors for the product space to be in the coupled or
uncoupled representations and many have difficulty in writing an operator in
a matrix form in the chosen basis. For example, when 26 students were asked
to choose a basis for two spin-1/2 particles and write down the matrix
corresponding to the operator $\hat {S}_1 \cdot \hat {S}_2 =(\hat {S}^2-\hat
{S}_1^2 -\hat {S}_2^2 )/2$ in that basis, 15{\%} of the students could not
find a complete set of basis vectors for the product space. Moreover, those
who chose the uncoupled representation often had difficulty figuring out how
to write $\hat {S}^2$ in a matrix form even though they were given the
appropriate Clebsch-Gordon Coefficient (CGC) table to write the coupled
states in terms of uncoupled states and vice-versa. About 33{\%} of the
students did not realize that the basis vectors in the coupled
representation are eigenstates of the operator $\hat {S}^2$, so that the
matrix elements of $\hat {S}_1 \cdot \hat {S}_2 $ can be calculated more
easily in the coupled representation than in the uncoupled representation.
Some students mistakenly thought that the basis vectors in the product space
are simply a collection of the basis vectors for the subspaces. For example,
for the two spin-1/2 particle system, about 8{\%} of the students
incorrectly wrote down the basis vectors as $\left| {s_1 =1/2,m_1 =1/2}
\right\rangle $, $\left| {s_1 =1/2,m_1 =-1/2} \right\rangle $, $\left| {s_2
=1/2,m_2 =1/2} \right\rangle $, and $\left| {s_2 =1/2,m_2 =-1/2}
\right\rangle $ and constructed incorrect $2\times 2$ matrices for the
operators they were asked to write in the matrix form in their chosen basis.
\vskip 0.5em
\noindent\textbf{\textit{Difficulty B.2: Incorrectly believing that if the operator
matrix is diagonal in one representation, it must also be diagonal in
another representation}}
\vskip 0.5em
To evaluate students understanding of operators in coupled and uncoupled
representations, the following multiple-choice question was given to 25
students.

\begin{enumerate}
\item[\textbullet] \textit{Suppose the Hamiltonian of a two spin-1/2 particle system is }$\hat {H}=\gamma ({\hat {\vec{ {S}_1}}} \cdot \vec {B}_1 +{\hat {\vec{ {S}_2}}} \cdot \vec {B}_2 )$\textit{ in which the magnetic fields }$\vec {B}_1 $\textit{ and }$\vec {B}_2 $\textit{ are both in the z-direction but with different magnitudes and $\gamma$ is a suitable constant. Choose all of the following statements that are correct.}
	\item[] \textit{(1) The Hamiltonian is a diagonal matrix in the coupled representation }$\left| {S^2,S_1^2, S_2^2, S_z } \right\rangle$.
	\item[] \textit{(2) The Hamiltonian is a diagonal matrix in the uncoupled representation }$\left| {S_1^2, S_{1z} } \right\rangle \otimes \left| {S_2^2, S_{2z} } \right\rangle$.
	\item[] \textit{(3) The Hamiltonian is a }$2\times 2$\textit{ matrix }$\hat {H}=\frac{\gamma \hbar B_1 }{2}\left( {{\begin{array}{*{20}c}
 1 \hfill & 0 \hfill \\
 0 \hfill & {-1} \hfill \\
\end{array} }} \right)+\frac{\gamma \hbar B_2 }{2}\left( {{\begin{array}{*{20}c}
 1 \hfill & 0 \hfill \\
 0 \hfill & {-1} \hfill \\
\end{array} }} \right)$\textit{in the uncoupled representation.}
\end{enumerate}

Since the basis vectors in the coupled representation, $\left| {S^2,S_1^2
,S_2^2, S_z } \right\rangle $, are not the eigenstates of the Hamiltonian,
some off-diagonal elements of $\hat {H}$ will be non-zero. These
off-diagonal matrix elements can be calculated by writing the coupled states
in terms of uncoupled states. However, when the $\hat {H}$ matrix is expressed in
the uncoupled representation, all of the off-diagonal elements are zero
since the basis vectors $\left| {S_1^2, S_{1z} } \right\rangle \otimes
\left| {S_2^2, S_{2z} } \right\rangle $ are the orthogonal eigenstates of
$\hat {H}$. While the correct answer is (2) only, about 40{\%} of the
students chose both the options (1) and (2). Some students incorrectly
believed that the Hamiltonian must be diagonal in both the coupled and
uncoupled representations. In individual discussions, students were asked to
write the operator $\hat {S}_{1z} +\frac{1}{2}\hat {S}_{2z} $ for a two
spin-1/2 particle system in matrix form in the product space. During
individual discussions, a student incorrectly believed that $\hat {S}_{1z}
+\frac{1}{2}\hat {S}_{2z} $ is diagonal in the coupled representation. When
he was told that the matrix was not diagonal in the coupled representation,
he claimed ``{\ldots}$\hat {S}_{1z} +\hat {S}_{2z} $ is a diagonal matrix in
the coupled basis. How can there be any difference between that operator and
the operator $\hat {S}_{1z} +\frac{1}{2}\hat {S}_{2z} $ when it is also a
superposition of $\hat {S}_{1z} $ and $\hat {S}_{2z} $?''. The student had
failed to observe that $\hat {S}_z=\hat {S}_{1z} +\hat {S}_{2z} $ is a very special
superposition of $\hat {S}_{1z} $ and $\hat {S}_{2z} $ which is diagonal in
both the coupled and uncoupled representations but other linear
superpositions of $\hat {S}_{1z} $ and $\hat {S}_{2z} $ will not be diagonal
in the coupled representation.

\subsection{Difficulty in constructing an operator matrix in the product space and realizing that the matrix corresponding to an operator could be very different in a different basis}

As described in section II, the matrix corresponding to an operator will
depend on the basis chosen. However, many students had difficulty realizing
that the matrix of the same operator can be very different for a different
basis. Below, we describe some related difficulties in more depth:
\vskip0.5em
\noindent\textbf{\textit{Difficulty C.1: Mistakenly adding the operators in different
Hilbert spaces algebraically to construct the operator for the product space
as if they act in the same Hilbert space}}
\vskip0.5em
We find that the students often have difficulty in constructing matrix
representations of operators correctly in the product space. For example,
when 26 students were asked to construct the matrix of $\hat {S}_{1z} +\hat
{S}_{2z} $ in a suitable basis, about 1/4 of them incorrectly claimed that
the resulting matrix is two dimensional and they simply added up the
matrices of the operators $\hat {S}_{1z} $ and $\hat {S}_{2z} $, i.e.,
\[
\hat {S}_{1z} +\hat {S}_{2z} =\frac{\hbar }{2}\left( {{\begin{array}{*{20}c}
 1 \hfill & 0 \hfill \\
 0 \hfill & {-1} \hfill \\
\end{array} }} \right)_1 +\frac{\hbar }{2}\left( {{\begin{array}{*{20}c}
 1 \hfill & 0 \hfill \\
 0 \hfill & {-1} \hfill \\
\end{array} }} \right)_2.
\]
Some of these students placed subscripts 1 and 2 on the matrices to
differentiate the two spin-1/2 particles, but most just merged them into a
single matrix. Similar difficulties were found when these students were
asked to construct a matrix for the operator $\hat {S}_1 \cdot \hat {S}_2 $
in any suitable basis of their choice. For example, three out of the 26
students simply multiplied the $2\times 2$ matrices corresponding to each of
the spins and expressed the result as another $2\times 2$ matrix. Also, more
than half of the students had difficulty realizing that the operator $\hat
{S}_{1z} $ in the product space of two spin-1/2 systems is a $4\times 4$
matrix (and not a $2\times 2$ matrix).
\vskip 0.5em
\noindent\textbf{\textit{Difficulty C.2: Incorrectly believing that the dimension of
the operator matrix depends on the choice of basis vectors}}
\vskip 0.5em
The dimension of the product space is independent of the basis or
representation chosen. For example, both the uncoupled and coupled
representations for the two spin-1/2 particle system have four basis vectors
since the product space is four-dimensional. Several students displayed an
inconsistency in interpreting the dimension of the product space depending
upon the basis chosen. For example, 5 out of 11 students incorrectly
believed that the matrix for the operator $\hat {S}_{1z} +\hat {S}_{2z} $ is
two-dimensional in the uncoupled representation. However, when explicitly
asked about the same operator in the coupled representation, they could
correctly construct a $4\times 4$ diagonal matrix with the eigenvalues of
$\hat {S}_{z}=\hat {S}_{1z} +\hat {S}_{2z} $ in the diagonal positions using the basis
vectors $\left| {1,1} \right\rangle $, $\left| {1,-1} \right\rangle $,
$\left| {1,0} \right\rangle $, and $\left| {0,0} \right\rangle $.
Discussions with individual students suggest that some of them may know that
the operator matrices are in general different in different basis sets but
they were unclear about the fact that the dimension of the product space
should always be equal to the number of linearly independent vectors in that
space and it cannot depend on the choice of basis vectors.
\vskip 0.5em
\noindent\textbf{\textit{Difficulty C.3: The Hamiltonian of the system must be known
in order to construct a matrix for an operator other than the Hamiltonian
operator}}
\vskip 0.5em
We asked the students to construct the matrix for the operator $\hat
{S}_{1z} +\frac{1}{2}\hat {S}_{2z} $ of a two spin-1/2 particle system with
only spin degrees of freedom involved. We found that some students believed
that the Hamiltonian of the system must be given in order for them to be
able to find the matrix elements of other operators. Discussions with
individual students suggest that this misconception originates from several
sources. First, some students believed that since the basis vectors are
often selected to be the eigenstates of the Hamiltonian, these are the only
basis vectors that can be used to construct the matrix for any operator.
Also, students were taught how to construct the Hamiltonian matrix for a
single electron spin in a uniform magnetic field (Larmor precession of spin)
and later they were taught how to construct the matrix of a Hamiltonian such
as $\hat {H}=\gamma (\hat {S}_1 \cdot \vec {B}+\hat {S}_2 \cdot \vec {B})$ in
the product space. There is also emphasis throughout the course on the role
of the Hamiltonian in determining the time evolution of the system and
allowed energies. Thus, some of the students over-generalized the importance
of the Hamiltonian in other contexts and claimed that they cannot construct
the matrix for the operator $\hat {S}_{1z} +\frac{1}{2}\hat {S}_{2z} $ in the
product space without knowing the Hamiltonian of the system.

\subsection{Difficulty in finding the probabilities for measuring an observable in a product space}

A particular choice of basis vectors for the product space may be suitable
for answering questions related to the probabilities of measuring a
particular observable. For example, if the question is related to the
probabilities of measuring the observable $S_{1z} $ or $S_{2z} $ in a
product state written in the coupled representation, it is convenient to go to
the uncoupled representation. Similarly, writing the states in the product
space in the coupled basis may be convenient for other probabilities, e.g.,
the probability of measuring the square of the magnitude of the total spin angular
momentum $\vec {S}^2$.

Students had difficulties in choosing a convenient basis in the product
space for answering questions related to probabilities of measuring a
particular observable. This difficulty was partly due to the fact that
students did not realize which basis vectors were eigenvectors of the
operator corresponding to a particular observable and that it is easy to
find the probabilities of measuring an observable if the state of the system
is written in terms of the eigenstates of that observable. Another type of
difficulty was related to transforming from one basis to another (e.g., from
coupled to uncoupled or vice versa) using the CGC table provided and
collecting all of the coefficients of a given state before taking the
absolute square of a coefficient to find the probability of measuring an
observable which has a definite value in that state. This latter difficulty
in collecting the coefficients of a particular state before taking the
absolute square to find the probability in product space is similar to those
observed for a single spin. It can be illustrated with the following
example. Suppose a single spin-1/2 state is given by the following
expression ($\left| {s,m_s } \right\rangle =\left| {1/2,\pm 1/2}
\right\rangle )$ after certain manipulations:
\begin{eqnarray*}
\left| \psi \right\rangle &=&\left( {a\left| {\frac{1}{2},\frac{1}{2}}
\right\rangle +{a}'\left| {\frac{1}{2},-\frac{1}{2}} \right\rangle }
\right)\\
&&+\left( {b\left| {\frac{1}{2},\frac{1}{2}} \right\rangle +{b}'\left|
{\frac{1}{2},-\frac{1}{2}} \right\rangle } \right),
\end{eqnarray*}
where $a$, ${a}'$, $b$ and ${b}'$ are constant coefficients. When we asked
17 students to calculate the probability of obtaining $\hbar /2$ for
measuring the observable $S_z $ (corresponding to $m_s =1/2)$, 12{\%} of the
students incorrectly responded that it was $\left| a \right|^2+\left| b
\right|^2$. However, the coefficients with the same basis vector $\left|
{s=1/2,m_s =1/2} \right\rangle $ should be combined first as $(a+b)\left|
{s=1/2,m_s =1/2} \right\rangle $ to yield the correct probability for
obtaining $\hbar /2$ to be $\left| {a+b} \right|^2$. This difficulty related
to finding the appropriate probability amplitude by combining the
coefficients of the same basis vectors persisted when dealing with a product
space and using the CGC table to transform from one basis to another. Such
difficulty may also partly be related to the general difficulties students
have in expressing quantities in the simplest algebraic form.

\subsection{Difficulty in dealing with the product space corresponding to the addition of orbital angular momentum and spin angular momentum of a single particle}

Some students thought that the addition of angular momentum formalism only
applies to a quantum system containing two or more particles (e.g., for
adding the spin angular momentum of different particles). They did not
realize that the orbital angular momentum and the spin angular momentum of a
single particle also follow the same rule of addition of angular momentum.
For example, since the orbital angular momentum state $\left| {\ell, m}
\right\rangle $ can be expressed in the form of a spherical harmonic
function $Y_\ell ^m $, the angular momentum part of a specific state of a
single spin-1/2 particle can be written as, e.g., $\sqrt {2/3} Y_1^0 \left|
{s=1/2,m_s =1/2} \right\rangle +\sqrt {1/3} Y_1^1 \left| {s=1/2,m_s =-1/2}
\right\rangle $ in the uncoupled representation. If we use the CGC table to
transform the basis vectors from the uncoupled representation to the coupled
representation, the same state can be written as $\left| {j=3/2,m_j =1/2}
\right\rangle $ where $j$ is the quantum number for the total angular
momentum and $m_j $ is the quantum number for the $z$-component of the total
angular momentum. However, when we asked students to find the outcome and
the corresponding probability if they measure the $z$-component of the total
angular momentum of a particle in the state $\sqrt {2/3} Y_1^0 \left|
{s=1/2,m_s =1/2} \right\rangle +\sqrt {1/3} Y_1^1 \left| {s=1/2,m_s =-1/2}
\right\rangle $, only 5 out of 11 students answered this question correctly.
Many students either forgot that the spherical harmonic function $Y_\ell ^m
$ corresponds to the orbital angular momentum state $\left| {\ell, m}
\right\rangle $ or mistakenly thought that $Y_1^0 $ and $Y_1^1 $ represent
the spin states $\left| {s=1/2,m_s =\pm 1/2} \right\rangle $. Interviews
with the students also indicated that some students have difficulties in
differentiating the related concepts about the orbital angular momentum and
the total angular momentum of a single particle. Interviewed students were
confused and were surprised to know that the addition of angular momentum
formalism is applicable regardless of whether we add orbital and spin
angular momenta, two orbital angular momenta, or two spin angular momenta.

\section{Improving students' understanding with the quantum interactive
learning tutorial (QuILT)}

We have developed a QuILT to help reduce the difficulties faced by students
in learning about the preliminaries of angular momentum addition. The QuILT
focuses on the preliminaries, e.g., the dimensionality of the product space,
how to choose a complete set of basis vectors for the product space, and how
to use these basis vectors to calculate the matrix elements for an operator
and construct an operator matrix. The QuILT focuses on helping students
learn about the coupled and uncoupled representations, but does not teach
them about going from one representation to another. This is because during
the investigation of difficulties we found that most students were
struggling with the preliminaries and those who learned the preliminaries
were able to comprehend the treatment in the common textbooks about how to
change basis. The QuILT builds on the prior knowledge of students as
determined by our investigation of difficulties. The concepts covered in the
QuILT are gradually built up in a guided inquiry-based approach \cite{21}. The
QuILT was developed based on the difficulties found by written surveys and
interviews. The development of the QuILT went through a cyclical interactive
process which included the following stages: (1) development of the
preliminary version based on a theoretical analysis of the underlying
knowledge structure and research on student difficulties; (2) implementation
and evaluation of the QuILT by administering it individually to students;
(3) determining its impact on student learning and assessing what
difficulties remained; (4) refinements and modifications based on the
feedback from the implementation and evaluation.

As noted earlier, the QuILT on the addition of angular momentum helps with
the issues related to finding the dimensionality of the product space and a
complete set of basis vectors in each of the coupled and uncoupled
representations. The QuILT also strives to help the students learn how to
use these basis vectors to construct any operator matrix in the coupled or
uncoupled representation by calculating the matrix elements (sandwiching the
operator between the state vectors). From the QuILT, students can learn that
an operator matrix may not be the same in the coupled and uncoupled
representations; regardless, the dimensionality of a product space in
coupled and uncoupled representations is always the same. Students are given
an opportunity to generalize all of the above issues from the case of two
spin-1/2 particles (4 dimensional product space) to other cases with higher
dimensions. However, the QuILT does not explicitly teach the students how to
write the basis vectors in coupled representations in terms of the basis
vectors in uncoupled representations or how to convert a matrix in the
coupled representation to a matrix in the uncoupled representation. As noted
earlier, many students were able to follow the treatment in the standard
textbooks to accomplish such tasks once they had worked on the QuILT to
learn the preliminaries. Such topics may be included in future supplementary
tutorials. In the current version of the QuILT, we focused on topics with
which students had great difficulties, including the dimensionality of a
product space, choosing a complete set of basis vectors for the product
space, the differences between the coupled and uncoupled representations,
constructing any operator matrix in a given basis, etc.

After iterating different versions of the QuILT with individual students
until the post-test performance of individual students who worked on the
QuILT significantly improved compared to the pre-test performance, it was
administered in undergraduate quantum mechanics classes after traditional
instruction on the addition of angular momentum. On average, students spend
around 1.5 hours on the QuILT on the addition of angular momentum, which
contains about 60 questions (most of which are in the multiple-choice format
to ensure that the QuILT can eventually be turned into a web-based tutorial
which students can use as a self-study tool and obtain appropriate feedback
if they click on an incorrect option for a multiple-choice question).

\subsection{Dimension of Hilbert space}

Before the QuILT on the basics related to the addition of angular momentum,
students are asked to work on a warm-up part which helps them with the
basics of a single spin system. Then, students work on the QuILT related to
the addition of angular momentum which has two parts, one part related to
the uncoupled representation and the other to the coupled representation. At
the beginning of the first part of the QuILT, students are asked about the
dimension of the product space for the two spin-1/2 particle system.
Together with the correct answer that the dimension is $4=2\times 2, $ a
distracter answer $4=2+2$ was also given in the multiple-choice question.
This strategy forces students to notice the difference between these two
answers and learn about the product space dimension by discussing their
answers with their peers. Note that the QuILT can be used as a self-study
tool, but when students work on it in class, peer discussion is exploited
throughout. Then the students go through a guided approach to construct
the basis vectors in the uncoupled representation for two spin-1/2 particles
(each with the $z$-component of spin quantum numbers $\pm 1/2)$,  e.g., $\left|
{1/2} \right\rangle _1 \otimes \left| {-1/2} \right\rangle _2 $ or $\left|
\uparrow \right\rangle _1 \otimes \left| \downarrow \right\rangle _2 $, and
learn about the fact that the operators $\hat {S}_{1z} $ and $\hat {S}_{2z}
$ only act on their respective subspaces in the uncoupled representation.
After this help in constructing basic understanding of the uncoupled
representation, students are asked the following multiple-choice question:

\begin{enumerate}
\item[\textbullet] \textit{Choose all the statements that are correct.}
\item[] \textit{(1)} $\left| {1/2} \right\rangle _1 \otimes \left| {1/2} \right\rangle _2 $ \textit{is an eigenstate of }$\hat {S}_{1z} $\textit{ and }$\hat {S}_{2z} $\textit{ but not }$\hat {S}_1^2 $\textit{ or }$\hat {S}_2^2. $
\item[] \textit{(2)} $\left| {1/2} \right\rangle _1 \otimes \left| {1/2} \right\rangle _2 $ \textit{is an eigenstate of }$\hat {S}_{1z}, \hat {S}_{2z}, \hat {S}_1^2 $\textit{ and }$\hat {S}_2^2. $
\item[] \textit{(3)} $\left| {1/2} \right\rangle _1 \otimes \left| {1/2} \right\rangle _2 $ \textit{is an eigenstate of }$\hat {S}_{1z}, \hat {S}_{2z}, \hat {S}_1 $\textit{ and }$\hat {S}_2. $
\end{enumerate}
Students discuss their answers with peers (correct answer is (2) only) and
learn that in the uncoupled representation, the basis vectors are
eigenstates of the individual spin operators $\hat {S}_{1z} $, $\hat {S}_{2z}
$, $\hat {S}_1^2 $ and $\hat {S}_2^2 $. They also learn to calculate the
individual matrix elements and construct various matrices in the uncoupled
representation (in the first part of the QuILT). For example, students learn
that $(\hat {S}_{1z} +\hat {S}_{2z} )\left| {1/2} \right\rangle _1 \otimes
\left| {-1/2} \right\rangle _2 =(\hbar /2-\hbar /2)\left| {1/2}
\right\rangle _1 \otimes \left| {-1/2} \right\rangle _2 =0$. For the
operator $\hat {S}_{1z} +\hat {S}_{2z} $, they are guided to conclude that
one of the matrix elements is $_2\langle-1/2|\otimes\, _1\langle1/2|(\hat {S}_{1z} +\hat {S}_{2z} )\left| {1/2}\right\rangle _1 \otimes \left| {-1/2} \right\rangle _2 =0$ and they also
practice how to find other matrix elements.

In order to generalize their understanding of the product space to more
complicated situations, students are later asked to consider the product
space of a three spin-1/2 particle system in the uncoupled basis. One
question explicitly asks them to consider the dimension in this case as
follows:

\begin{itemize}
\item \textit{What is the dimensionality of the spin space of a three spin-1/2 system?}
\item[] \textit{(a)} $2$ \textit{ (b)} $2+2+2=6$ \textit{ (c)} $3^2=9$ \textit{ (d)} $2^3=8$
\end{itemize}
Here students are given an opportunity to think about and discuss with peers
the fact that the dimension of a product space is the product of the
dimensions of the subspaces (and hence the correct answer is (d)). They are
further asked to construct a complete set of eight basis vectors and then
calculate several diagonal and off-diagonal matrix elements of the operator
$\hat {S}_{1z} +\hat {S}_{2z} +\hat {S}_{3z} $ in the uncoupled
representation.

\subsection{Constructing matrices for different operators for the product
space of two spin-1/2 systems in the uncoupled representation}

In the QuILT, students are asked to calculate the matrices of the following
operators in the uncoupled representation given that a uniform magnetic field $\vec B$ is pointing in the $+z$ direction: $\hat {H}_1 =(4E_0 /\hbar )({\hat
{\vec {{S}_1}}} \cdot {\hat {\vec {{S}_2}}} )$ and $\hat {H}_2 =-\mu ({\hat {\vec{{S}_1}}} \cdot \vec
{B}+{\hat{\vec{ {S}_2}}} \cdot \vec {B})$. They are told that $\mu $
and $4E_0 /\hbar $ are constants. Students must contemplate the properties of the
operators $\hat {S}_{1z} +\hat {S}_{2z} $ and ${\hat {\vec {{S}_1}}} \cdot {\hat {\vec{ {S}_2}}}
$. Since the basis vectors in the uncoupled representation are orthonormal
eigenstates of the operators $\hat {S}_{1z} $ and $\hat {S}_{2z} $, all the
off-diagonal matrix elements of the operator $\hat {S}_{1z} +\hat {S}_{2z} $
are zero. After helping students learn about how to construct the matrix for
the operator $\hat {S}_{1z} +\hat {S}_{2z} $, students learn how to construct the matrix
for a more
complicated operator, ${\hat {\vec{ {S}_1}}} \cdot {\hat{\vec{ {S}_2}}} $, in the uncoupled representation. They are first asked the
following question to help them think about why it is more convenient to
write the operator in the form $\hat {S}_{1x} \hat {S}_{2x} +\hat
{S}_{1y}  \hat {S}_{2y} +\hat {S}_{1z}  \hat {S}_{2z} $ rather
than $\hat {S}^2-\hat {S}_1^2 -\hat {S}_2^2 $ to calculate the matrix
elements in the uncoupled basis (without access to the CGC table).

\begin{itemize}
\item \textit{Consider the following conversation between Pria and Mira:}
\item[] \textbf{\textit{Pria}}\textit{: Is }$(4E_0 /\hbar ){\hat {\vec {{S}_1}}} \cdot {\hat {\vec { {S}_2}}} $\textit{ or }$(2E_0
/\hbar )(\hat {S}^2-\hat {S}_1^2 -\hat {S}_2^2 )$\textit{ the more convenient form for writing }$\hat {H}_1 $\textit{ in matrix form in the uncoupled representation?}
\item[]
\textbf{\textit{Mira}}\textit{: Since the basis vectors }$\left| {m_s } \right\rangle _1 \otimes \left| {m_s }
\right\rangle _2 $\textit{ are not the eigenstates of }${\hat {\vec{ {S}_1 }}} \cdot {\hat {\vec{ {S}_2}}} $ or $\hat {S}^2-\hat {S}_1^2
-\hat {S}_2^2 $\textit{, we have to be careful. It is the form }$(4E_0 /\hbar ){\hat {\vec{ {S}_1}}} \cdot {\hat {\vec{ {S}_2}}} $\textit{ that is more useful because we can write }${\hat {\vec{ {S}_1}}} \cdot
{\hat {\vec{{S}_2}}} =\hat {S}_{1x}  \hat {S}_{2x} +\hat {S}_{1y}  \hat
{S}_{2y} +\hat {S}_{1z}  \hat {S}_{2z} $\textit{. Then we can write the x and y components of spin in terms of the raising and lowering operators and we know how they act on }$\left| {m_s } \right\rangle _1
\otimes \left| {m_s } \right\rangle _2$.
\item[] \textit{Do you agree with Mira? Explain.}
\end{itemize}

The students learn to rewrite the operator ${\hat {\vec{{S}_1}}} \cdot {\hat{\vec{ {S}_2}}} $
using the raising and lowering operators such that ${\hat {\vec { {S}_1}}} \cdot {\hat {\vec{
{S}_2}}} =(\hat {S}_{1-}  \hat {S}_{2+} +\hat {S}_{1+}  \hat {S}_{2-}
)/2+\hat {S}_{1z} \hat {S}_{2z} $ and they practice applying the
raising and lowering operators to the basis vectors in the uncoupled
representation as in the following example (with correct answer (1)):

\begin{enumerate}
\item[\textbullet] \textit{Which one of the following equations is correct?}
	\item[] \textit{(1)} $\hat {S}_{1-} \hat {S}_{2+} \left| {1/2} \right\rangle _1 \otimes \left| {1/2} \right\rangle _2 =(\hat {S}_{1-} \left| {1/2} \right\rangle _1 )\otimes (\hat {S}_{2+} \left| {1/2} \right\rangle _2 )=0$
	\item[] \textit{(2)}  $\hat {S}_{1-} \hat {S}_{2+} \left| {1/2} \right\rangle _1 \otimes \left| {1/2} \right\rangle _2 =(\hat {S}_{1-} \left| {1/2} \right\rangle _1 )\otimes (\hat {S}_{2+} \left| {1/2} \right\rangle _2 )=\hbar ^2\left| {1/2} \right\rangle _1 \otimes \left| {1/2} \right\rangle _2 $
	\item[] \textit{(3)}  $\hat {S}_{1-} \hat {S}_{2+} \left| {1/2} \right\rangle _1 \otimes \left| {1/2} \right\rangle _2 =(\hat {S}_{1-} \left| {1/2} \right\rangle _1 )\otimes (\hat {S}_{2+} \left| {1/2} \right\rangle _2 )=2\hbar ^2\left| {1/2} \right\rangle _1 \otimes \left| {1/2} \right\rangle _2 $
\end{enumerate}
Students learn that calculating the matrix elements of an operator in the
uncoupled basis can be achieved by expressing the operator as a combination
of $\hat {S}_{1z} $, $\hat {S}_{2z} $, $\hat {S}_{1\pm } $, and $\hat
{S}_{2\pm } $ since operators corresponding to each spin-1/2 system act on its own subspace. Students
are asked to explain the characteristics of the operators that will be
diagonal in the uncoupled representation and they are given multiple
opportunities to test what they predict in concrete situations and reconcile
the differences between their predictions and observations if there are any.
At the end of the first part of the QuILT, students are given the following
question to help them understand that the basis vectors can be chosen
according to convenience.

\begin{itemize}
\item \textit{Consider the following conversation between Andy and Caroline and explain with whom you agree.}
\item[] \textbf{\textit{Andy}}\textit{: For the question about choosing a basis for two spin-1/2 systems, we do not necessarily have to choose a basis in the product space which consists of eigenstates of }$\hat {S}_{1z} $\textit{ and }$\hat {S}_{2z}$.
\item[] \textbf{\textit{Caroline}}\textit{: I disagree. We must choose a basis in the product space such that the basis vectors are eigenstates of }$\hat {S}_{1z} $\textit{ and }$\hat {S}_{2z}$.
\end{itemize}
The QuILT helps students learn, e.g., that usually if an operator can be
represented by a diagonal matrix in a particular basis, that basis is more
convenient than others. It is not necessary to always choose a basis in
the product space which is an eigenstate of $\hat {S}_{1z} $ and $\hat
{S}_{2z} $. Discussion about these questions also leads to a smooth
transition to other basis vectors, e.g., from the uncoupled to the coupled
representation.

\subsection{Introducing the coupled representation}

The section of the QuILT that introduces the coupled representation asks
students to list all of the possible total spin quantum numbers $s$ for the
total spin angular momentum $\vec {S}=\vec {S}_1 +\vec {S}_2 $ for the
product space of two spin-1/2 systems. They also list the quantum numbers
for the $z$-component of total spin $m_s $ when $s=1$ and 0. Students contemplate
why a complete set of coupled states denoted by quantum numbers $s$ and $m_s $
and written as $\left| {s,m_s } \right\rangle $ forms a set of basis vectors
for the product space for a system of two spin-1/2 systems. Some guided
inquiry-based questions help students learn to apply different operators
such as $\hat {S}^2$, $\hat {S}_1^2 $, $\hat {S}_2^2 $ and $\hat {S}_z $ on
the states $\left| {s,m_s } \right\rangle $. Students also verify that the
basis vectors in the coupled representation are orthonormal. As shown in the
multiple choice question below, the QuILT also helps students contemplate
the differences between the coupled and uncoupled basis vectors.

\begin{enumerate}
\item[\textbullet] \textit{Choose all of the following statements that are correct about the differences between the ``coupled'' and ``uncoupled'' representations of the multi-spin system.}
\item[] \textit{(1) Working entirely within the coupled representation, you cannot decompose the product state of a two-spin system into products of states of each individual spin.}
\item[] \textit{(2) Working entirely within the uncoupled representation, you can decompose the product state of a two-spin system into products of states of each individual spin.}
\item[] \textit{(3) The basis vectors in the uncoupled representation are eigenstates of }$\hat {S}_1^2,  \hat {S}_{1z},  \hat {S}_2^2 $\textit{, and }$\hat {S}_{2z} $\textit{, whereas the basis vectors in the coupled representation are eigenstates of }$\hat {S}^2, \hat {S}_1^2,  \hat {S}_2^2 $\textit{, and }$\hat {S}_z =\hat {S}_{1z} +\hat {S}_{2z}$.
\end{enumerate}
All of the three options in the question above are correct. Through these
types of questions, students learn that in the coupled representation the
basis vectors in the product space are such that the individual states of
the two particles cannot always be separated from each other. However, the basis
vectors of the coupled representation can be written as linear combinations
of the basis vectors of the uncoupled representation. They also observe that
the basis vectors in the coupled and uncoupled representations are not the
eigenstates of the same operators. For example, the basis vectors in the
coupled representation are the eigenstates of the square of the total spin
operator $\hat {S}^2$, but the basis vectors in the uncoupled representation
are not the eigenstates of this operator. On the other hand, the activities
in the QuILT help them learn that the basis vectors in both the coupled and
uncoupled representations are eigenstates of the operators $\hat {S}_1^2 $
and $\hat {S}_2^2 $.

\subsection{Constructing the matrices for various operators in the product
space of two spin-1/2 systems in the coupled representation}

In the second part of the QuILT, students are given the task of writing the
same operators $\hat {H}_1 =\gamma ({\hat {\vec {{S}_1}}} \cdot {\hat {\vec{ {S}_2}}} )$ and $\hat
{H}_2 =\mu ({\hat {\vec{ {S}_1}}} \cdot \vec {B}+{\hat {\vec{ {S}_2}}} \cdot \vec {B})$ in matrix
form in the coupled representation that they had earlier learned to write in
the uncoupled representation via a guided inquiry process (without using the
CGC Table). They are asked to compare the matrices in the coupled
representation with those in the uncoupled representation. They also learn
that in the coupled representation, it is convenient to write the operator
${\hat {\vec{ {S}_1}}} \cdot {\hat{\vec{ {S}_2}}} =(\hat {S}^2-\hat {S}_1^2 -\hat {S}_2^2 )/2$ so
the matrix elements can be easily calculated. There are discussions in the
QuILT to help students understand why an operator, e.g., ${\hat {\vec{ {S}_1}}} \cdot
{\hat {\vec { {S}_2}}} $, is diagonal in the coupled basis but non-diagonal in the
uncoupled basis. Students also explore and learn that the operator $\hat
{H}_2 =\mu ({\hat {\vec{ {S}_1}}} \cdot \vec {B}+{\hat {\vec{ {S}_2}}} \cdot \vec {B})$ is
diagonal in both the coupled and uncoupled representations since the basis
vectors in the coupled representation are the eigenstates of the operator
$\hat {S}_z =\hat {S}_{1z} +\hat {S}_{2z} $ (since the magnetic field is in the $z$ direction). Then they are asked to express
the matrix for $\hat {H}_2 $ in the block diagonal form where all the
non-zero terms are confined to a smaller block rather than being spread out
in the full $4\times 4$ matrix. This process helps students understand that
they can arrange the order of basis vectors as they wish while constructing
the matrix.

\section{Preliminary evaluation}

Based on the investigation of students' difficulties, we designed a pre-test
and a post-test to assess the issues related to the addition of angular
momentum. The pre-test was administered to the entire undergraduate quantum
mechanics classes (altogether 26 students) in the 2009 spring semester and
2010 spring semester after traditional instruction. After the pre-test, the
same two groups of students worked on the QuILT and the post-test was
administered to them the following week in class. The questions in the
pre-test and the post-test were similar, but used product spaces for quantum
systems with different spin. In particular, in the pre-test, the system
contained two spin-1/2 particles, while the system in the post-test had one
spin-1/2 particle and one spin-1 particle. The pre-test question asks:

\begin{itemize}
\item \textit{Two spin-1/2 systems (with the spin quantum numbers }$s_1 =1/2$\textit{ and }$s_2 =1/2$\textit{) at fixed locations in space (only consider spin degrees of freedom) interact with each other, and with a uniform magnetic field }$\vec {B}$\textit{ pointing in the $+z$ direction. When the magnetic field is off, the interaction between the spins is given by the Hamiltonian }
\[
\hat {H}_1 =(4E_0 /\hbar ){\hat {\vec {{S}_1}}} \cdot {\hat {\vec { {S}_2}}} =(2E_0 /\hbar
)(\hat S^2-\hat S_1^2 -\hat S_2^2 ),
\]
\textit{where }${\hat {\vec{ {S}}}}={\hat {\vec{ {S}_1}}} +{\hat {\vec{ {S}_2}}} $\textit{ and }$E_0 $\textit{ is a constant. The magnetic field interacts with each spin as follows:}
\[
\hat {H}_2 =-\mu ({\hat {\vec {{S}_1}}} \cdot \vec {B}+{\hat {\vec{ {S}_2}}} \cdot \vec {B}).
\]
\item[] \textit{(a) Write down a complete set of basis vectors for the vector space of a system of two spin-1/2 particles. Explain the labels you are using to identify your basis states.}
\item[] \textit{(b) Express the Hamiltonian }$\hat {H}_1 $\textit{ in the basis you have chosen. (Write it down as an }$N\times N$\textit{ matrix).}
\item[] \textit{(c) Express the Hamiltonian }$\hat {H}_2 $\textit{ in the basis you have chosen.}
\item[] \textit{(d) Are both }$\hat {H}_1 $\textit{ and }$\hat {H}_2 $\textit{ diagonal matrices in the basis you chose?}
\end{itemize}

Questions (a), (b) and (c)  weigh 3 points each and question (d) weighs 1
point. The correctness percentage of all the 26 students answering question
(a) in the pre-test and post-test is listed in table 1. In the pre-test, 9
students chose the uncoupled basis and 4 students wrote the coupled basis
vectors correctly. Another 7 students correctly expressed the coupled
singlet/triplet states using the uncoupled basis vectors, e.g., $\left|
{1,1} \right\rangle =\left| {\uparrow \uparrow } \right\rangle $ and $\left|
{0,0} \right\rangle =\frac{1}{\sqrt 2 }\left( {\left| {\downarrow \uparrow }
\right\rangle -\left| {\uparrow \downarrow } \right\rangle } \right)$. Other
students gave incorrect responses, e.g., $\left| {1/2,\pm 1/2} \right\rangle
_1 $ and $\left| {1/2,\pm 1/2} \right\rangle _2 $, which indicated that they
believed that the basis vectors in the product space are the same as the
basis vectors in the subspaces. In the post-test, most of the students chose
the coupled basis vectors when answering the first question (but for the
product space of a spin-1/2 and a spin-1 system). However, four students did
not provide the correct answer because they mistakenly treated the post-test
question as a two spin-1/2 particles system or a three spin-1/2 particles
system.

The correctness percentages of the students answering questions (b) and (c)
in the pre-test and post-test are listed in table 2. Only two out of the
twenty-six students correctly calculated the matrices for both the operators
$\hat {H}_1 $ and $\hat {H}_2 $ in the pre-test. About 27{\%} of the
students incorrectly calculated the matrices for the operators $\hat {H}_1 $
and $\hat {H}_2 $ in the four dimensional Hilbert space by simply adding the
matrices for the spin operators in the two dimensional subspaces. Other
students had no idea about how to calculate the matrix elements of the
operators for a given set of basis vectors. In the post-test, 21 out of the
26 students correctly knew that the matrix elements can be obtained by
sandwiching the operator with the corresponding basis vectors. Among these
21 students, 66{\%} of them correctly calculated the matrices of both $\hat
{H}_1 $ and $\hat {H}_2 $ using the coupled representation. Some students
who selected either the coupled or uncoupled representation had difficulties
in calculating the matrix elements, especially for some off-diagonal
elements in $\hat {H}_1 $.
\begin{table}
\caption{Correctness percentage of 26 students answering question
(a) in the pre-test and post-test. The italicized items represent the types
of students' correct or incorrect answers and the corresponding numbers
stand for the number of students giving a particular type of answer. The
item ``mixed'' means that the students expressed the coupled singlet/triplet
states using the uncoupled basis vectors. The item ``dimension'' represents
the mistakes related to the dimension of the Hilbert space, e.g., only
writing down the basis vectors in the subspace of one of the spins.}
\begin{ruledtabular}
\begin{tabular}{llllllll}
\multicolumn{4}{c}{Pre-test Question (a)} & \multicolumn{4}{c}{Post-test Question (a)}  \\
\multicolumn{2}{c}{Correct 77{\%}} & \multicolumn{2}{c}{Incorrect 23{\%}} & \multicolumn{2}{c}{Correct 85{\%}} & \multicolumn{2}{c}{Incorrect 15{\%}}  \\
\colrule
\textit{Uncoupled}& 9& \textit{Dimension}& 4& \textit{Uncoupled}& 6& \textit{Dimension}& 4 \\
\textit{Coupled}& 4& \textit{Mixed}& 2& \textit{Coupled}& 16& \textit{Mixed}& 0 \\
\textit{Mixed}& 7& & & \textit{Mixed}& 0& &  \\
\end{tabular}
\end{ruledtabular}
\end{table}

\begin{table*}
\caption{Correctness percentages of 26 students answering questions
(b) and (c) in the pre-test and post-test. The item ``sandwich'' means that
the students calculated the matrix elements by sandwiching the operator with
the basis vectors. The item ``simply add'' represents the mistake of
calculating the matrices by simply adding the matrices for the spin
operators in different subspaces as though they were in the same vector
space. The number corresponding to each item stands for the number of
students giving a particular type of answer.}
\begin{ruledtabular}
\begin{tabular}{lllllllll}
& \multicolumn{4}{c}{Pre-test Questions (b) and (c)} & \multicolumn{4}{c}{Post-test Questions (b) and (c)}  \\
\colrule
\raisebox{-4.50ex}[0cm][0cm]{b}& \multicolumn{2}{c}{Correct 8{\%}} & \multicolumn{2}{c}{Incorrect 92{\%}} & \multicolumn{2}{c}{Correct 54{\%}} & \multicolumn{2}{c}{Incorrect 46{\%}}  \\
 &\textit{Sandwich}&2&\textit{Sandwich}&0&\textit{Sandwich}&14&\textit{Sandwich}&7 \\
&&&\textit{Simply Add}&7&&&\textit{Simply Add}&0 \\
 &&&\textit{No Idea}&17&&&\textit{No Idea}&5 \\
\colrule
\raisebox{-4.50ex}[0cm][0cm]{c}& \multicolumn{2}{c}{Correct 8{\%}} & \multicolumn{2}{c}{Incorrect 92{\%}} & \multicolumn{2}{c}{Correct 73{\%}} & \multicolumn{2}{c}{Incorrect 27{\%}}  \\
 & \textit{Sandwich}& 2& \textit{Sandwich}& 0& \textit{Sandwich}& 19& \textit{Sandwich}& 2 \\
 & & & \textit{Simply Add}& 7& & & \textit{Simply Add}& 0 \\
 & & & \textit{No Idea}& 17& & & \textit{No Idea}& 5 \\
\end{tabular}
\end{ruledtabular}
\end{table*}

The correctness percentages of the students answering question (d) in the
pre-test and post-test are listed in table 3. As shown in table 3, only 8
out of the 26 students answered question (d) correctly in the pre-test while
the other students did not know whether the operator matrices should be
diagonal or not in a particular basis. In the post-test, 85{\%} of the
students who had appropriately chosen either the coupled or uncoupled
representation in question (a) correctly answered question (d). The average
correctness percentage for all the questions in the pre-test was 40{\%} in
2009 (with 9 students) and 27{\%} in 2010 (with 17 students). After the
students learned the topic using the QuILT for the addition of angular
momentum, their post-test average score increased to 73{\%} and 72{\%} in
2009 and 2010, respectively.
\begin{table}
\caption{Correctness percentage of 26 students answering question
(d) in the pre-test and post-test. Students who did not provide the correct
response either provided wrong answers or left the question blank (which are
represented by the items ``wrong'' and ``no answer''). The number
corresponding to each item stands for the number of students giving a
particular type of answer.}
\begin{ruledtabular}
\begin{tabular}{llllllll}
\multicolumn{4}{c}{Pre-test Question (d)} & \multicolumn{4}{c}{Post-test Question (d)}  \\
\multicolumn{2}{c}{Correct 31{\%}} & \multicolumn{2}{c}{Incorrect 69{\%}} & \multicolumn{2}{c}{Correct 85{\%}} & \multicolumn{2}{c}{Incorrect 15{\%}}  \\
\colrule
\textit{Right }& 8& \textit{Wrong}& 9& \textit{Right }& 22& \textit{Wrong}& 2 \\
\textit{Answer}& & & & \textit{Answer}&& &  \\
& & \textit{No Answer}& 9& & & \textit{No Answer}& 2 \\
\end{tabular}
\end{ruledtabular}
\end{table}

We have conducted a Wilcoxon signed-rank test for students' pre-test and
post-test total scores and the result indicated significant improvement in
the post-test performance ($z$-value $>$ 3.5). We also conducted a two-tailed
t-test on students' scores for each question and their total scores in the
pre- and post-tests. The p-values are less than 0.001 for all questions
except for question (a) (on which students had performed reasonably well in
the pretest). While the post-test scores are statistically significantly
better than the pre-test scores ($p$-value $<$ 0.001), students' understanding
can be improved further. One central reason for the low score on the
post-test was that many students treated the post-test problem as though it
was a product space of two spin-1/2 systems as opposed to a product of one
spin-1/2 and one spin-1 system. However, the fact that a greater percentage
of students chose the coupled representation in the post-test than in the
pre-test indicates that after using the QuILT, the students are aware of the
fact that choosing the basis vectors to be the eigenstates of the operators
can make the calculations easier. Also, when answering questions (b) and (c)
in the post-test, most students who had used the tutorial can construct the
diagonal or off-diagonal operator matrix by sandwiching the operator with
the basis vectors in the product space.

\section{Summary and Conclusion}

We find that students have many common difficulties related to the addition
of angular momentum. For example, many students were unclear about the
dimension of the product space and they incorrectly believed that the
dimension of the product space is the sum of the dimensions of the
subspaces. Students also had difficulty in distinguishing between the basis
vectors in the coupled and uncoupled representations and had difficulty in
determining how to choose an appropriate basis for the product space to
answer questions related to the measurements of different observables. While
changing basis using a CGC table, students also had difficulty in
determining how to calculate the probabilities of measuring different
observables. Students struggled to construct the matrix of an operator in a
convenient basis in the product space. Some students believed that the
dimension of a product space in the coupled and uncoupled representations is
different. In particular, some students simply added the matrices for two
spin-1/2 particles to construct the matrix of the operator $\hat {S}_{1z}
+\hat {S}_{2z} $ such that the resulting matrix in the product space was
still two dimensional. Some students had difficulty understanding why the
operator $\hat {S}_{1z} +\hat {S}_{2z} $ is diagonal in the coupled
representation but $\hat {S}_{1z} +\frac{1}{2}\hat {S}_{2z} $ is not. Some
believed that they must be given the Hamiltonian of the system in order to
write any operator in the matrix form in a given basis.

We developed a research-based QuILT to improve students' understanding of
the addition of angular momentum. It provides a guided approach to bridge
the gap between the quantitative and conceptual issues related to addition
of angular momentum and helps students connect different concepts and build
a knowledge structure. The QuILT keeps students actively engaged in the
learning process. Our preliminary assessment shows that the QuILT improves
students' understanding of concepts related to the addition of angular
momentum.

\begin{acknowledgments}
This material is based upon work supported by the National Science
Foundation under Grant Numbers NSF-PHY 1202909 and 0855424. We thank the
professors who administered the free-response and multiple-choice questions
to their students during the investigation of difficulties and administered
the QuILT.
\end{acknowledgments}


\begin{thebibliography}{99}
\bibitem{1}D. Styer, Common misconceptions regarding quantum mechanics, Am. J.
Phys. \textbf{64}, 31 (1996).
\bibitem{2}P. Jolly, D. Zollman, S. Rebello, and A. Dimitrova, Visualizing
potential energy diagrams, Am. J. Phys. \textbf{66}, 57 (1998).
\bibitem{3}I. D. Johnston, K. Crawford, and P.R. Fletcher, Student difficulties
in learning quantum mechanics, Int. J. Sci. Educ. \textbf{20}, 427
(1998).
\bibitem{4}D. Zollman, S. Rebello, and K. Hogg, Quantum physics for everyone:
Hands-on activities integrated with technology, Am. J. Phys. \textbf{70},
252 (2002).
\bibitem{5}M. Wittmann, R. Steinberg, E. Redish, Investigating student
understanding of quantum physics: spontaneous models of conductivity, Am.
J. Phys. \textbf{70}, 218 (2002); Research on Teaching and Learning of Quantum Mechanics,
papers presented at the National Association for Research in Science Teaching,Boston, MA, 1999,
<perg.phys.ksu.edu/papers/narst/>; Also Am. J. Phys.
70(3), (2002) published in conjunction with Gordon Conference on
physics research and education on quantum mechanics.
\bibitem{6}C. Singh, Student understanding of quantum mechanics at the beginning
of graduate instruction, Am. J. Phys. \textbf{76}, 277 (2008); C. Singh,
Student Understanding of Quantum Mechanics, Am. J. Phys., \textbf{69}, 885,
(2001);
\bibitem{7}L. D. Carr and S. B. McKagan, Graduate quantum mechanics reform, Am.
J. Phys. \textbf{77}, 208 (2009).
\bibitem{8}C. Singh, M. Belloni, and W. Christian, Improving student's
understanding of quantum mechanics, Physics Today \textbf{8}, 43 (2006).
\bibitem{9}A. J. Mason and C. Singh, Do advanced students learn from their mistakes
without explicit intervention?, Am. J. Phys., \textbf{78}, 760, (2010); A.
Mason and C. Singh, Reflection and self-monitoring in Quantum Mechanics, AIP
Conf. Proc. \textbf{1179}, 197 (2009).
\bibitem{10}G. Zhu and C. Singh, Surveying students' understanding of quantum
mechanics, AIP Conf. Proc. \textbf{1289}, 301 (2010); C. Singh and G. Zhu,
Students' understanding of the addition of angular momentum, AIP Conf. Proc.
\textbf{1413}, 355 (2012); G. Zhu and C. Singh, Students' difficulties with
quantum measurement, AIP Conf. Proc. \textbf{1413}, 387 (2012). C. Singh and G. Zhu,
Cognitive Issues in Learning Advanced Physics: An Example from Quantum
Mechanics, AIP Conf. Proc. \textbf{1179}, 63 (2009);
S. Siddiqui and C. Singh, Surveying instructors' attitudes and approaches to teaching quantum mechanics,
AIP Conf. Proc. \textbf{1289}, 297 (2010).
\bibitem{11}G. Zhu and C. Singh, Surveying students' understanding of quantum
mechanics in one spatial dimension, Am. J. Phys., \textbf{80}, 252 (2012).
\bibitem{12}C. Singh, Interactive learning tutorials on quantum mechanics, Am. J.
Phys., \textbf{76}, 400 (2008); C. Singh, Helping Students Learn Quantum
Mechanics for Quantum Computing, AIP Conf. Proc. \textbf{883}, 42, (2007).
\bibitem{13}G. Zhu and C. Singh, Improving students' understanding of quantum
measurement. I. Investigation of difficulties, Phys. Rev. ST Phys. Educ.
Res. \textbf{8}, 010117 (2012); G. Zhu and C. Singh, Improving students'
understanding of quantum measurement. II. Development of research-based
learning tools, Phys. Rev. ST Phys. Educ. Res. \textbf{8}, 010118 (2012).
\bibitem{14}Improving Students' Understanding of Quantum Mechanics, G. Zhu, Ph.D.
Dissertation, University of Pittsburgh, 2011.
\bibitem{15}E. Mazur, Peer Instruction: A User's Manual. (Prentice Hall, Upper
Saddle River, NJ, 1997).
\bibitem{16}C. Singh, Transfer of Learning in Quantum Mechanics, AIP Conf.
Proc. \textbf{790}, 23 (2005).
\bibitem{17}S. Y. Lin and C. Singh, Categorization of quantum mechanics problems
by professors and students, Euro. J. Phys. 31, 57 (2010); S. Y. Lin and
C. Singh, Assessing Expertise in Quantum Mechanics using Categorization
Task, AIP Conf. Proc. \textbf{1179}, 185 (2009).
\bibitem{18}C. Singh, Student difficulties with quantum mechanics formalism,
AIP Conf. Proc. \textbf{883}, 185 (2007).
\bibitem{19}C. Singh, Assessing and improving student understanding of quantum
mechanics, AIP Conf. Proc. \textbf{818}, 69 (2006).
\bibitem{20}F. W. Byron and R. W. Fuller, Mathematics of Classical and Quantum
Physics. (Addison-Wesley, Reading, MA, 1969).
\bibitem{21}G. Zhu and C. Singh, Improving students' understanding of quantum
mechanics via the Stern-Gerlach experiment, Am. J. Phys. \textbf{79},
499 (2011); G. Zhu and C. Singh, Students' Understanding of
Stern-Gerlach Experiment, AIP Conf. Proc. \textbf{1179}, 309 (2009).

\end{thebibliography}

\end{document}